\documentclass[11pt]{amsart}
\usepackage{geometry}                
\usepackage{graphicx}
\usepackage{amssymb,amsmath}
\usepackage[english]{babel}
\usepackage[ansinew]{inputenc}
\usepackage{textcomp} 
\usepackage{flafter}  
\usepackage{natbib} 
\usepackage{bm}  

\begin{document}
\newcommand{\const}{\mathrm{const}}
\newcommand{\mD}{\mathcal{D}}
\newcommand{\omegaP}{\omega^{\prime}}
\newcommand{\reff}[1]{(\ref{#1})}
\newcommand{\eref}[1]{Eq.\reff{#1}}
\title{Viscoresistive MHD Configurations of Plasma in Accretion Disks}
%


\author[Montani and Benini]{Giovanni Montani$^{1234\dag}$ \hspace{5mm} {\tiny and} \hspace{5mm} Riccardo Benini$^{34\ddag}$\\[5mm]
\\
             \tiny{$^{1}$ENEA - C.R. Frascati (Department F.P.N.) 
              Via Enrico Fermi, 45 (00044), Frascati (Rome), Italy\\[5mm]
              $^{2}$ICRANet - C.C. Pescara, P. della Repubblica, 10 (65100), Pescara, Italy\\
              $^{3}$Department of Physics (G9) 
              ``Sapienza'' Universit\`a di Roma,\\ 
              Piazzale A. Moro, 5 (00185), Rome, Italy\\
               $^{4}$ICRA - International Centre for Relativistic Astrophysics\\
              $^{\dag}${Montani@icra.it}\\
              $^{\ddag}${riccardo.benini@icra.it}}
}

\maketitle

\begin{abstract}
We present a discussion of two-dimensional magneto-hydrodynamics (MHD) configurations, concerning the equilibria of accretion disks of a strongly magnetized astrophysical object. We set up a viscoresistive scenario which generalizes previous two-dimensional analyses by reconciling the ideal MHD coupling of the vertical and the radial equilibria within the disk with the standard mechanism of the angular momentum transport, relying on dissipative properties of the plasma configuration. 

The linear features of the considered model are analytically developed and the non-linear configuration problem is addressed, by fixing the entire disk profile at the same order of approximation. Indeed, the azimuthal and electron force balance equations are no longer automatically satisfied when poloidal currents and matter fluxes are included in the problem. These additional components of the equilibrium configuration induce a different morphology of the magnetic flux surface, with respect to the ideal and simply rotating disk.\\
\end{abstract}

\section{Preliminaries}

Understanding the mechanism of accretion that compact objects show
in the presence of lower dense companions,
is a long standing problem in astrophysics \citep{B01}.
In fact, while in absence of a significant magnetic
field of the massive body the disk configuration
is properly described by the fluidodynamical
approach \citep{S73}, the situation becomes a bit puzzling when
we deal with a strongly magnetized source.
Since over the last three decades,
increasing interest raised about the mechanism
of the angular momentum transport across the disk
profile. The solution of this problem stands
as settled down in the form of a standard model,
at least in the case of small (intrinsic) magnetic fields.
This context prevents the formation of high magnetic
back-reactions inside the plasma disk and the gas-like
approximation is predictive.
We remark how
it is  known from the observations \citep{verbunt82}
that the accretion profile takes place often
with the morphology of a thin disk configuration.
In this limit of a thin gaseous disk,
the hydrodynamical equilibria underlying the
accretion process are cast into a one-dimensional paradigm,
(see among the first analyses of the problem
\citep{pringlerees72,S73,lyndenpringle74}).
In such a fluidodynamical scenario, the accretion
mechanism relies on the angular momentum transfer
as allowed by the shear viscosity properties of the
disk material. The differential angular rotation
of the radial layers is associated with a
non-zero viscosity coefficient,
accounting for diffusion and turbulence phenomena.
Indeed, microscopic estimations of the viscosity
parameter indicate that the friction of different
disk layers would be unable to maintain a sufficiently
high accretion rate, and therefore
non-linear turbulent features of the equilibrium are inferred.

However, when the magnetic field $\vec{B}$ of the central object
is strong enough, the electromagnetic
back-reaction
of the disk plasma becomes relevant \citep{RuffiniWilson}.
As shown in \cite{C05}, the Lorentz force induces
a coupling between the radial and the vertical
equilibria, which deeply alters the local morphology
of the system. In particular, the radial dependence of
the disk profile acquires
an oscillating
character, modulating the background structure too.
The existence of such a coupling
breaks down the one-dimensional nature of the
problem and suggests a
revision of the original point of view
of the standard model.

Furthermore, in \cite{CR06} it is discussed 
how the plasmas, characterised by a $\beta$-parameter (i.e. the ratio of the thermostatic pressure $p$ to the magnetic one $B^{2}/8\pi$)
close to unity, exhibit an oscillating-like
mass density and the
disk is decomposed in a ring profile.
Thus, the analyses in \cite{C05,CR06} demonstrate that
the details of the disk equilibria are relevant in
establishing a crystalline local structure inside
the disk.
This ideal MHD result constitutes an opposite
point of view with respect to the idea of a diffusive
magnetic field within the disk, as discussed in \cite{B01}.
The striking interest in the details of  this local
disk morphology relies on the idea that
jets of matter and radiation are emitted by virtue
of the strong magnetic field and of the axial symmetry of
the accretion profiles 
(see for instance \cite{L96}).

Such an ideal two-dimensional MHD analysis is pursued
by neglecting the accretion rate of the disk in the leading
order (see also the global analysis presented in \cite{Ogilvie97}) and thus without facing the question
about the azimuthal balance of the forces acting
on the electrons. In fact, in these works \citep{Ogilvie97,C05,CR06},
toroidal currents (and toroidal matter fluxes)
are addressed only.
We analyze in some detail the implications that
accounting for the azimuthal and electron force balance
equilibrium equations has on the local profile
of the disk. We pursue the setting up of these
two additional equations in the same approximation
limit of the analysis in \cite{C05} and in \cite{CR06} and we
arrive to establish a direct relation between the
radial infall velocity and the azimuthal Lorentz force.
This result relies on addressing a link between the
turbulent viscosity coefficient and the resistivity
one. In particular, in the limiting case $\beta\to\infty$, we
are able to fix the vertical dependence
of the plasma configuration,
differently from the issue in \cite{C05},
for the same limiting case.
The radial velocity we derive this way has an
oscillating behavior in its radial dependence and therefore
has a zero net radial average. This
matter infall profile is able to provide a local
non-zero accretion rate and therefore it reconciles
the crystal structure, outlined in \cite{C05},
with a real accretion feature active in the disk.

The paper is organized as follows.
In subsection \ref{sec:2} we describe the basic
equations of the stationary MHD. In subsection \ref{sec:3}
the standard hydrodynamic theory of accretion
disk is presented. In Section \ref{sec:4} the bidimensional accretion
disk is discussed in the framework of MHD. Section
\ref{sec:azimuthaleq} is devoted to an analysis of
the azimuthal equation and on the balance of the
force acting on the electrons along the tangential
direction. The linear approximation of this
bidimensional model is discussed in Section
\ref{sec:linearapprox}, while the non linear configuration is developed in Section \ref{sec:nonlinearapprox}. Concluding Remarks follow

\subsection{MHD of the Steady State} \label{sec:2}

The behavior of a plasma embedded into a magnetic
and a gravitational fields can be described
by a MHD approach for a wide range of configurations.
In the case of hydrogen-like ions and when a steady
state holds,
the system configuration is governed
by the following set of equations
(as written in Cartesian coordinates, and 
we note that repeated indices are intended as summed).
\begin{subequations}
\begin{eqnarray}
\label{threee}
&\partial_i \left( \epsilon v_i\right) = 0\\
&\epsilon v_l\partial _lv_i = -\partial _ip
+\partial _l \left[\mathcal{D} \left( \partial _iv_l +
\partial _lv_i - \displaystyle\frac{2}{3}\delta _{il}\partial _kv_k
\right) \right] 
-\partial _i\chi + \displaystyle\frac{1}{c}\epsilon_{ijk}
J_jB_k\\
&v_l\partial _lT + \displaystyle\frac{2}{3}T\partial _lv_l = 0
\, , 
\end{eqnarray}
\end{subequations}
where
the mass density $\epsilon$ and 
the total (electron plus ion) pressure $p$
are related via the temperature $T$,
according to the equation of state
$p = 2 K T \epsilon /m$, where $K$ denotes the Boltzmann
constant and $m$ is the proton mass. Here 
$\mathcal{D}$ denotes the shear viscosity coefficient,
$J_i$ the density current vector, $v_{i}$ the plasma velocity field,
while, $\chi$ stands for the Newton potential due to the
central body of mass $M_\mathrm{S}$
(the self-gravity of the plasma being negligible), i.e.
$\chi(x^{i}) = G M_\mathrm{S}/\sqrt{\delta_{ij} x^{i} x^{j}}$.
Instead, the plasma acquires
an internal electromagnetic structure whose
back-reaction can be relevant.

The behavior of the total electric field $E_i$ and
magnetic field $B_i$ is described
by the Maxwell equations standing in Gaussian units as
\begin{subequations}\label{Maxeq}
\begin{eqnarray}
&\epsilon _{ijk}\partial _jB_k =
\displaystyle\frac{4\pi }{c}J_i&\label{Maxeq1}\\
&\partial _lE_l = 4\pi \rho&\\
&\epsilon _{ijk}\partial _jE_k = 0
\quad \Rightarrow \quad  E_i = -\partial _i \Phi&\\
&\partial _lB_l = 0 
\quad \Rightarrow \quad
B_{i} = \epsilon _{ijk}\partial _jA_k &
\, ,
\end{eqnarray}
\end{subequations}
where $A_i$ denotes the potential vector,
$\rho$ the electric charge density, $\Phi$ the electric potential and the
electric and magnetic fields are related by the
MHD condition
\begin{equation}
E_l = -\displaystyle\frac{1}{c}\epsilon _{lmn}v_mB_n
\, .
\label{mhdcond}
\end{equation}
Recalling the relation
\begin{equation}
\label{epsp}
\epsilon _{ijk}\epsilon_{ilm} =
\delta _{jl}\delta _{km} -
\delta _{jm}\delta _{kl}
\end{equation}
and by means of  \eref{Maxeq1}
the Lorentz force rewrites as
\begin{equation}
\displaystyle\frac{1}{c}\epsilon _{ijk}J_jB_k =
\displaystyle\frac{1}{4\pi}\left( - B_l\partial _iB_l +
B_l\partial _lB_i\right)
\, .
\label{Lf}
\end{equation}
The scheme we traced above completely fixes the
steady plasma configuration, once assigned the central object 
morphology.

\subsection{The Standard Model of Accretion disks\label{sec:3}}

The characterization of an accretion disk is
obtained by addressing an axisymmetric MHD configuration
(described by cylindrical coordinates
$r,\,\phi ,\,z$)
for the plasma surrounding the central object.
However, some fundamental features of the accretion
process are fixed by the fluidodynamical approach
describing the matter infall through
a gas disk profile, as outlined in \cite{S73}.
We aim to fix how the viscoresistive
approach is not well-grounded
on a microscopical level and a new perspective
is required, as based on typical features observed in laboratory plasmas \citep{C05,CR06}.

In the standard model of accretion,  the configuration of an axisymmetric thin disk
is determined by the fluidodynamical equilibria,
which take place within the central gravitational field.

The thin character of the disk depth allows us to
simplify the configuration problem, by integrating over
the vertical profile, so fixing an effective one-dimensional
hydrodynamical problem.
When averaged out of the vertical direction \citep{B01},
the radial equilibrium reduces to the condition that
the angular velocity $\omega$ of the disk takes the Keplerian profile
\begin{equation}
\omega (r)=\omega _\mathrm{K}=\sqrt{\frac{GM_\mathrm{S}}{r^3}}
\,.
\label{kepom}
\end{equation}
This statement is equivalent to neglect the
role played in the equilibrium by the radial
pressure gradient, retaining the centripetal force
exerted by the central object as the dominant effect.
This Keplerian nature of the disk is well-grounded and 
significant deviations from such a behavior are expected
in advective dominated regimes only.

The vertical equilibrium corresponds to 
the gravothermal configuration confining the disk profile,
and it is therefore
governed by the equation
\begin{equation}
\frac{dp}{dz} + \omega^2_\mathrm{K}z\epsilon = 0 
\, ,
\label{verteq}
\end{equation}
which, for an isothermal disk of temperature $T$
and equation of state $p = v_\mathrm{s}^2\epsilon$
(where $v_\mathrm{s}^2=2K_\mathrm{B}T/m$ is the sound velocity), gives
the exponential decay of the mass density over the
equatorial plane value $\epsilon_0(r)$, i.e.
\begin{equation}
D(z^2)\equiv \frac{\epsilon }{\epsilon _0} =
\exp\left(- \displaystyle\frac{z^2}{H^2}\right)
\, .
\label{dzquad}
\end{equation}
Here, $H = \sqrt{2v_\mathrm{s}^2/\omega_\mathrm{K}^2}$ provides
an estimation for the real 
half-depth of the disk, which, in principle
would have an infinite vertical extension.

The azimuthal equilibrium describes the angular
momentum transport across the disk, by virtue of
a viscous stress tensor component $\tau _{r\phi}$
which enters the
relation (first integral of the azimuthal equation)
\begin{equation}
\dot{M}_\mathrm{d} (L - L_\mathrm{d}) = -2\pi r^2\tau _{r\phi}
\quad , \quad \tau _{r\phi} =
\mathcal{D} r^2\frac{d\omega}{dr}
\, .
\label{azzeqq}
\end{equation}
Here $\mathcal{D}$ must be thought  as 
a turbulent viscosity coefficient,
while $L$ is the angular momentum per unit mass.
Moreover, 
$L_\mathrm{d}$ is a fixed value of the specific angular
momentum and
\begin{equation}
\dot{M}_\mathrm{d} \equiv -2\pi r\int _{-H}^{H}(\epsilon v_r)dz = 
-2\pi r\sigma v_r
\, ,
\label{accrate}
\end{equation}
is the mass accretion
rate, associated to the radial velocity $v_r<0$ and to the
surface mass density $\sigma \equiv \int_{-H}^{H}\epsilon dz$.
Finally the continuity equation
\begin{equation}
\frac{1}{r}\frac{d(r\epsilon v_r)}{dr} +
\frac{d(\epsilon v_z)}{dz} = 0
\, ,
\label{conteq}
\end{equation}
once integrated over the vertical direction provides
with the fundamental relation
$\dot{M}_\mathrm{d} = \const > 0$.

Recalling that the model is dominated by the Keplerian
feature $\omega \simeq \omega_\mathrm{K}$ and observing
that $d\omega_\mathrm{K}/dr = -3\omega_\mathrm{K}/2r$, we
arrive to the following form for the angular
momentum transport versus the plasma turbulent viscosity 
\begin{equation}
\dot{M}_\mathrm{d}(L - L_\mathrm{d}) = 3\pi \mathcal{D}\omega_\mathrm{K}r^2
\, .
\label{accrelf}
\end{equation}

\paragraph{Plasma viscosity}  
This picture of the accretion mechanism is however affected
by a discrepancy between theory and observations of
real systems. In fact, 
the viscosity coefficient $\mD$ 
in the disk is too small
if estimated by the microscopic plasma (or atomic)
structure. Its microscopic features are unable
to account for the accretion rates observed
in some astrophysical systems, like X-ray binaries.
The observed accretion rates,
as estimated by the increasing disk luminosity 
${\dot{\mathcal{L}}}_\mathrm{d}\sim GM_\mathrm{S}{\dot{M}}_\mathrm{d}/r_\mathrm{S}$ ($r_\mathrm{S}$ being the radius of the central object), 
would require a larger value of $\mathcal{D}$,
(see the discussion in \cite{S73}).
The solution to this discrepancy is inferred in
the turbulent behavior arising in the disk when the 
Reynold number is sufficiently large.
Such turbulence in the disk plasma would then be 
responsible for the appearance of the large
value of the viscosity coefficient.

Since, by definition
$L=\omega r^2$, we can infer
\begin{equation}
\label{shakura}
\mD = 2\sigma v_\mathrm{t}H/3\,,
\end{equation} 
$v_\mathrm{t}$ being a turbulence
velocity, given by $v_\mathrm{t}=\alpha v_\mathrm{s}$,
where $\alpha$ is a free parameter.
It must be noted that
the axisymmetric disk is linearly stable with
respect to small perturbations preserving its symmetry and
the angular momentum conservation.
For a discussion of the onset of turbulence by MHD instabilities,
see the analysis presented in \cite{B98}.
Such a review work is
based on the Velikhov approach of \citeyear{V59}
and it requires non-linear interaction among perturbations
of very small amplitude.

\subsection{Comparison with the literature}

The difficulty of this turbulent scenario,
led B. Coppi to provide in \cite{C05} a local 
2-dimensional MHD formulation of the disk profile,
which argues how the notion of plasma turbulence
could be replaced by microscopic ring-like structures.
In this framework, that we address below,
the effect of turbulent viscosity has to be replaced
by fundamental plasma instabilities, observed in
laboratory experiments, like the so-called
\emph{Resistive Ballooning Modes}.
An interesting ideal two-dimensional MHD analysis
of a rotating disk equilibrium has been provided
by G. I. Ogilvie
in \cite{Ogilvie97}, where the thin disk configuration
is addressed in both the weak and strong magnetic field
limit (also solutions are derived
assuming self-similarity in the radial profile
of the non-thin case).
The hypotheses at the ground of
this work are the same as in \cite{C05,CR06}, but
the analyzed regimes are significantly different.
In fact, in \cite{Ogilvie97} the radial dependence
of the disk profile has a power-law structure and
the existence of the ring morphology does not emerge
for a strongly magnetized thin disk.
The main reason of such significant deviation between
these two ideal MHD approaches relies both on the
local character of the analysis of  Coppi (compared
to the global profile of Ogilvie)
and on the nature of the asymptotic expansion of
Ogilvie, which privileges the second
vertical derivatives with respect to the radial ones
(so altering the Laplacian term appearing in the
local equilibria fixed by Coppi).

Our analysis differs from both these two ideal
formulations of the axisymmetric MHD equilibria,
not only for including dissipative effects
(like viscosity and resistivity), but, overall
in view of considering poloidal currents and
matter fluxes. In fact in order to describe the accretion
phenomenon it is necessary to deal with radially
in-falling material and expectedly a non-zero
azimuthal Lorentz force.
These two features are not present in the approaches followed
in \cite{Ogilvie97,C05,CR06} and our
generalization involves dissipative effects too
in order to compare the vertical and radial
coupling, within the disk configuration, to
the paradigm of the standard approaches.

Indeed, the standard model treatment of the accretion process
is properly reviewed by \cite{B01}, where the
main features of the disk morphology and the main
successful issues are summarized in some detail.
For a more recent analysis of the questions
concerning the diffusive magnetic field living in
the plasma configuration and its possible enhancement
to get jet configuration see \cite{Sp08}. 
The models presented in these two reviews differ
deeply from the analyses by Coppi or by Ogilvie,
because the validity of the viscoresistive MHD approach is postulated
on a phenomenological ground, almost disregarding
the coupling between the radial and the vertical equilibria.

The main merit of our approach is in reconciling
the two-dimensional MHD scenario, including
microstructures of the plasma configuration, with the
the viscoresistive formulation of the standard theory.
Despite we are here not able to solve the resistivity
puzzle (see the dedicated paragraph of section \ref{electronforcebalance}),
we demonstrate that the microstructures determined by
B. Coppi for the ideal MHD approach still survive in
the presence of non-zero viscosity and resistivity
coefficients and of the related radial and vertical
matter velocity components. Demonstrating such a structural
stability of the crystalline profile, we open the
perspective to unify the microscopical and observational
points of view, expectedly on the level of plasma instabilities.
Anyway, the non-zero local accretion rate predicted
by our model represents a significant improvement
of the analysis in \cite{C05,CR06}, because it
upgrades the crystalline picture toward a realistic
disk configuration.

\section{Two-Dimensional MHD Model for an Accretion Disk \label{sec:4}}

Let us now specify the steady MHD theory,
discussed in Section 2, 
to the specific case of an accretion disk
configuration
around the compact (few units of Solar mass) and
strongly magnetized (a dipole-like field of
about $10^{12} Gauss$), i.e. a typical pulsar source.
The gravitational potential of the pulsar has the form
\begin{equation}
\chi (r\, ,z) = \frac{GM_\mathrm{S}}{\sqrt{ r^2 + z^2}}
\, ,
\label{Gravpot}
\end{equation}
It is worth noting that
the axial symmetry prevents any dependence
on the azimuthal angle $\phi$ of all the quantities involved in
the problem. In this respect the continuity equation
takes the explicit form (\ref{conteq}), 
which provides the following 
matter flux associated with the disk
morphology 
\begin{equation}
\epsilon \vec{v} =
-\frac{1}{r}\partial _z\Theta \vec{e}_r +
\epsilon \omega(r\, ,z^2)r\vec{e}_{\phi } +
\frac{1}{r}\partial _r\Theta \vec{e}_z
\, ,
\label{solconteq}
\end{equation}
where $\Theta (r\, , z)$ is an odd function of $z$ in order
to deal with a non-zero accretion rate, i.e.
\begin{equation}
\dot{M}_\mathrm{d} = -2\pi r\int _{-H}^{H}\epsilon v_rdz =
4\pi \Theta (r\, , H) > 0
\, ,
\label{Mdot}
\end{equation}

The magnetic field, characterizing the central
object, takes the form 
\begin{equation}
\vec{B} = -\frac{1}{r}\partial _z\psi \vec{e}_r +
\frac{I}{r}\vec{e}_{\phi } + 
\frac{1}{r}\partial _r\psi \vec{e}_z
\, ,
\label{vectorb}
\end{equation}
with $\psi = \psi (r\, ,z^2)$ and
$I = I(\psi \, , z)$. The similarity of the magnetic field and
matter flux structure, is due to their common divergence-less
nature.

We now develop a local model of the equilibrium,
as settled down around a radius value $r = r_0$,
in order to analytically investigate the effects induced on
the disk profile by the electromagnetic reaction of the plasma.
To this end we split the mass density and the pressure
contributions as
$\epsilon = \bar{\epsilon }(r_0,\, z^2) + \hat{\epsilon}$
and 
$p = \bar{p}(r_0,\, z^2) + \hat{p}$, respectively. 
The same way, we express the magnetic surface function
in the form 
$\psi = \psi _0(r_0) +
\psi _1(r_0\, , r-r_0\, ,z^2)$, with $\psi _1\ll \psi _0$.
The  quantities $\hat{\epsilon}$, $\hat{p}$ and $\psi _1$ describe the change induced by the currents which rise
within the disk imbedded into the external magnetic field
of the central object. In general these corrections are
small in amplitude but with a very short scale of variation.
Thus, we are lead to address the ''drift ordering'' for
the behavior of the gradient amplitude, i.e.
the first order gradients of the perturbations are of
zero-order, while the second order ones dominate.

As ensured by the corotation theorem \cite{Ferraro}, the angular
frequency of the disk rotation has to be expressed
via the flux function as $\omega(\psi )$. 
As a consequence, in the present split scheme, 
we can take the decomposition 
$\omega = \omega_\mathrm{K} +
\omega^{\prime }_0\psi_1$,
where $\omega_\mathrm{K}$ is the Keplerian term and
$\omega^{\prime }_0 = \const.$. This form for $\omega$
holds locally, as far as $(r - r_0)$ remains a sufficiently
small quantity, so that the dominant deviation 
from the Keplerian contribution is due to $\psi _1$.

Accordingly to the drift ordering, the profile of the toroidal
currents rising in the disk, has the form
\begin{equation}
J_{\phi } \simeq -\frac{c}{4\pi r_0}
\left(\partial ^2_r\psi _1 + \partial_z^2\psi _1\right)
\, .
\label{jphi}
\end{equation}
On the other hand, the azimuthal component of the
Lorentz force is related to the existence of the
function $I(\psi ,z)$ and it can be written as 
\begin{equation}
F_{\phi } \simeq \frac{1}{4\pi r_0^2}
\left(\partial _zI\partial _r\psi -
\partial_rI\partial _z\psi\right)
\, .
\label{fphi}
\end{equation}

\subsection{Vertical and Radial Equilibria \label{sec:verticalradialeq}}

We now fix the equations governing the vertical
and the radial equilibrium of the disk, distinguishing
the fluid components from the 
electromagnetic back-reaction (as develop by \cite{C05,CR06}).
Such a splitting of the MHD equations for the
vertical force balance gives
\begin{equation}
D(z^2) \equiv \frac{\bar{\epsilon}}{\epsilon _0(r_0)} =
\exp\left(-\frac{z^2}{H^2}\right)
\, , \hspace{5mm} \epsilon _0(r_0) \equiv \epsilon (r_0,\, 0)
\, , \hspace{5mm} H^2 \equiv \frac{4K_\mathrm{B}\bar{T}}{m\omega _\mathrm{K}^2}\;,
\end{equation}
\begin{equation}\label{verticalequilibrium}
\partial _z\hat{p} + \omega ^2_\mathrm{K}z\hat{\epsilon}
- \frac{1}{4\pi r_0^2}\left(
\partial ^2_z \psi_1 + \partial^2_r\psi _1\right)
\partial _z\psi_1 = 0 
\, ,
\end{equation}
Here, as in Section \ref{sec:3}, the behavior of the function $D(z^2)$
accounts for the pure thermostatic equilibrium
standing in the disk when the vertical gravity
(i.e. the Keplerian rotation) is sufficiently high
to provide a confined thin configuration,
while the temperature $T$ admits the representation
\begin{equation}
2 K_\mathrm{B} T \equiv m\frac{p}{\epsilon} =
m\frac{\bar{p} + \hat{p}}{\bar{\epsilon} +
\hat{\epsilon}} \equiv 2 K_\mathrm{B} (\bar{T} + \hat{T})
\, .
\label{temptot}
\end{equation}
The radial equation underlying the equilibrium
of the rotating layers of the disk, can be decomposed
into the dominant character of the Keplerian angular
velocity plus an equation describing the
behavior of the deviation
$\delta \omega$, i. e.
\begin{equation}
\omega \simeq \omega _\mathrm{K} + \delta \omega \simeq
\omega _0(\psi _0) + \omega_{0}^{\prime}\psi _1\;,
\end{equation}
\begin{equation}
\begin{split}\label{radialequilibrium}
2\omega _\mathrm{K}r_0(\bar{\epsilon} + \hat{\epsilon})
\omega _0^{\prime }\psi _1 &+
\frac{1}{4\pi r_0^2}\left(
\partial ^2_z \psi_1 + \partial^2_r\psi _1\right)
\partial _r\psi_1 =\\
&=\partial _r\left[
\hat{p} + \frac{1}{8\pi r_0^2}
\left(\partial_r\psi_1\right)^2\right]
+ \frac{1}{4\pi r_0^2}\partial_r\psi _1 \partial^2_z\psi_1
\end{split}
\end{equation}
Here we neglected  the presence of the poloidal currents,
associated with the azimuthal component of the magnetic
field.

Let us define the dimensionless functions 
$Y$, $\hat{D}$ and $\hat{P}$, in place of
$\psi _1$, $\hat{\epsilon}$ and $\hat{p}$, i.e.
\begin{equation}
Y\equiv \frac{k_0\psi _1}{\partial _{r_0}\psi _0}
\, , \hspace{5mm}
\hat{D}\equiv \frac{\beta \hat{\epsilon}}{\epsilon _0}
\, , \hspace{5mm}
\hat{P}\equiv \beta \frac{\hat{p}}{p_0}
\, ,
\label{deff}
\end{equation}
where $p_0\equiv 2K_\mathrm{B}\bar{T}\epsilon _0$
and $\beta \equiv 8\pi p_0/B^2_{0z} = 
1/(3\epsilon _z^2) \equiv k_0^2H^2/3$.
Here we introduced the fundamental wavenumber $k_0$
of the radial equilibrium, defined as
$k_0\equiv 3\omega _\mathrm{K}^2/v_\mathrm{A}^2$, with
$v_\mathrm{A}^2\equiv B^2_{z0}/4\pi \epsilon _0$, recalling
that $B_{z0} = B_{z}(r, z=0) = \partial _{r_0}\psi _0/r_0$.
Thus, we introduce the dimensionless
radial variable $x\equiv k_0(r - r_0)$, while
we assume that the fundamental length in the vertical
direction be $\Delta \equiv \sqrt{\epsilon _z}H$,
leading to define $u\equiv z/\Delta$.
By these definitions, the vertical equilibrium
can be restated as
\begin{equation}
\partial _{u^2}\hat{P} + \epsilon _z\hat{D}
- 2\left(\partial ^2_{x}Y +
\epsilon _z\partial ^2_{u}Y\right)
\partial _{u^2}Y = 0
\, ,
\label{vertad}
\end{equation}
while the radial configuration is fixed by
\begin{equation}
\left(D + \frac{1}{\beta }\hat{D}\right) Y + 
\left(\partial ^2_{x}Y +
\epsilon _z\partial ^2_{u}Y\right)(1 + \partial_xY )
+\frac{1}{2}\partial _x\hat{P} 
= 0 
\, .
\label{radad}
\end{equation}
The two equations above provide a coupled system for $\hat{P}$ and $Y$ once the quantities $D$ and $\hat{D}$ are assigned;  hence we are able to determine the disk
configuration induced by the toroidal currents.
However, it remains to be described the accretion features of
of the disk and the angular momentum transport across
the crystalline structure lacking in the original analysis developed in \citet{C05,CR06}.

\section{Azimuthal equation \label{sec:azimuthaleq}}

The equilibrium along the toroidal symmetry of the disk
is described by an equation 
which takes the exact expression
\begin{equation}
\epsilon v_r\partial _r(\omega r) +
\epsilon v_z\partial _z(\omega r) +
\epsilon \omega v_r = \frac{1}{r^2}\partial _r\left(
\mD r^3\partial _r\omega \right) +
\partial _z\left[ \mD \partial _z(\omega r) \right] 
+ F_{\phi }
\, , 
\label{exazeq}
\end{equation}
$F_{\phi }$ being the $\phi $-component of the
Lorentz force.
The corotation theorem, stating   $\omega = \omega (\psi )$, 
allows us to rewrite the equation above as
\begin{equation}
\begin{split}
\label{exazeq2}
&\epsilon rv_r\partial _r\psi +
\epsilon rv_z\partial _z\psi +
2\epsilon v_r \frac{\omega }{\omegaP }
= \\
&\frac{1}{r^2\omegaP}\partial _r\left(
\mD r^3
\omegaP
\partial _r\psi \right) +
\frac{1}{\omegaP}
\partial _z\left[ \mD
r\omegaP
\partial _z\psi \right]
+ \frac{F_{\phi }}{\omegaP }
\, ,
\end{split}
\end{equation}
($\omegaP=d\omega/d\psi$).
By virtue of the magnetic field form, we restate
the l.h.s. of this equation in the form
\begin{equation}
\begin{split}
\label{exazeqre}
&\epsilon r^2\left( v_rB_z - v_zB_r\right) +
2\epsilon v_r \frac{\omega }{\omegaP}
= \\
&\frac{1}{r^2\omegaP}\partial _r\left(
\mD  r^3
\omegaP
\partial _r\psi \right) +
\frac{1}{\omegaP}
\partial _z\left[ \mD
r\omegaP
\partial _z\psi \right]
+ \frac{F_{\phi }}{\omegaP}
\, . 
\end{split}
\end{equation}
In the local model we are addressing, by virtue of \eref{fphi}
the azimuthal equation stands at $r_0$ as 
\begin{equation}
\begin{split}
\epsilon r_0^2\left( v_rB_z - v_zB_r\right) +
2\epsilon v_r \frac{\omega _\mathrm{K}}{\omega ^{\prime }_0}
&= r_0 \mD_{0}
\left(\partial ^2_r\psi _1 + \partial_z^2\psi _1\right)
+\\&+ \frac{1}{4\pi r_0^2\omega ^{\prime }_0}
\left[\partial _zI\left( \partial _{r_0}\psi_0
+ \partial _r\psi _1\right) 
 - \partial_rI\partial _z\psi_1\right]
\, ,
\label{exazeqlo}
\end{split}
\end{equation}
where $\mD_{0} =\mD(r_{0})$.
This equation accounts for the angular momentum transport
across the disk, by relaying on the presence of a
viscous feature of the differential rotation in the spirit of Section 3.
Furthermore, in this scheme, we link the radial and vertical
velocity fields with the poloidal currents present in the
configuration. Indeed we aim to get an equilibrium picture
in which the disk accretion is induced by the
poloidal currents directly, i.e. we want to establish a
relation between $v_r$ and $F_{\phi}$. To this end,
as well as for the model consistence, we have to analyze
the structure of the electron force balance equation.

\subsection{The electron force balance equation}\label{electronforcebalance}

In the presence of a non-zero resistivity coefficient
$\eta$, the equation accounting for the electron force
balance, reads as
\begin{equation}
\vec{E} + \frac{\vec{v}}{c}\wedge \vec{B} =
\eta \vec{J}
\, . 
\label{efb1}
\end{equation}
Since the contribution of the resistive term is expected to be relatively small,
it turns out as appreciable only in the azimuthal component
of the equation above.
Thus, the balance of the Lorentz force has 
dominant radial and vertical components, 
providing the electric field in the form
predicted by the corotation theorem, i.e.
\begin{equation}
\vec{E} = -\frac{\vec{v}}{c}\wedge \vec{B} =
-\frac{d\Phi}{d\psi}\vec{\nabla }\psi =
-\frac{\omega }{c}\left( \partial _r\psi \vec{e}_r
+ \partial _z\psi \vec{e}_z\right)
\, .
\label{MHDeqcond}
\end{equation}
Since the axial symmetry requires
$E_{\phi}\equiv 0$,
the $\phi$-component of the equation above
takes the form 
\begin{equation}
v_zB_r - v_rB_z = c\eta J_{\phi }
\, .
\label{efb1x}
\end{equation}
In the local formulation around $r_0$,
the azimuthal component stands as follows 
\begin{equation}
v_rB_z - v_zB_r = 
\frac{\eta c^2}{4\pi r_0}
\left(\partial ^2_r\psi _1 + \partial_z^2\psi _1\right)
\, .
\label{efbloc}
\end{equation}
We now observe that,
substituting \eref{efbloc} in \eref{exazeqlo}, we get 
\begin{equation}
\begin{split}
2\epsilon v_r \frac{\omega _\mathrm{K}}{\omega ^{\prime }_0}
=& r_0\left( \mD_{0} - \frac{c^2\eta \epsilon}{4\pi}\right) 
\left(\partial ^2_r\psi _1 + \partial_z^2\psi _1\right)
\\
&+ \frac{1}{4\pi r_0^2\omega ^{\prime }_0}
\left[\partial _zI\left( \partial _{r_0}\psi_0
+ \partial _r\psi _1\right) 
 - \partial_rI\partial _z\psi_1\right]
\, .
\label{exazeqlocomb}
\end{split}\end{equation}
This relation together with the electron force balance 
one (\ref{efbloc}) provide a system for the two unknowns
$\Theta$ and $I$, when $\psi_1$ (i.e. $Y$) and $\epsilon$
(i.e. $D$ and $\hat{D}$) 
are given by the vertical (\ref{vertad}) and radial (\ref{radad}) equilibria. 

\paragraph{The resistivity puzzle}
Before proceeding with our analysis, it is worth
noting that the explanation for a high value
of the resistivity coefficient of the disk plasma
is necessary to make consistent the accretion scenario
of the standard model, but it is hard to be provided
and we are led to speak of a real resistivity
puzzle. In fact, in the limit of a small electromagnetic
back-reaction of the plasma, we deal with a negligible
vertical dependence of the flux surface function
(i.e. $B_r\sim 0$) and the azimuthal component
of the electron force balance \eref{efb1}
reduces to the simple form
$v_rB_z = -\eta J_{\phi}$. Since in the considered
limit, the toroidal current can not be significantly
intense, the only way to ensure a sufficiently high
radial velocity (able to account relevant accretion rates),
consists of accounting on the role played by
resistivity in fixing the equilibrium.
However, like the viscosity coefficient,
the quantity $\eta$ results to be very small if
microscopically estimated and the question arises
about the origin of strong resistive features of the disk.
In the standard model, an ''anomalous'' resistivity
coefficient is postulated in view of the turbulent
behavior associated to the Velikhov instability
of the rotating configuration. This scenario privileges
a link between the viscosity and resistivity parameters,
leading to values of order unity for
magnetic Prandtl number. In the study we pursue below,
despite the equilibrium is fixed for generic
values of these dissipative coefficients, the
linear and non linear solutions are derived under 
the assumptions of important viscoresistive
effects. Our aim here is not to solve the puzzle
of resistivity, but reconciling the crystalline
profile with the presence of poloidal currents
and matter fluxes. For a proposal of the mechanism
able to restore accretion in agreement to very small
values of the resistivity coefficient, see \cite{MGXII}.

\section{The linear approximation \label{sec:linearapprox}}

Let us first analyze the linear model, corresponding to
the request
$\partial _{r_0}\psi _0 \gg \partial _r\psi _1$ and 
$\partial _z\psi _1\simeq 0$. These conditions are
equivalent to impose on the dimensionless vertical (\ref{vertad}) and radial (\ref{radad}) equations, the restriction $Y\ll 1$ and to
approach the limit $\epsilon _z < 1$
(i.e. $\beta > 1$). Furthermore  the
pressure gradient is neglected because it
is not expected to be the responsible for a significant
deviation from the Keplerian disk.
Such an approximation will hold in the non-linear
and low $\beta$ values limits too.

In the linear regime, we clearly have to require
$\epsilon = \bar{\epsilon } + \hat{\epsilon}
\simeq \bar{\epsilon} = \epsilon _0 (r_0)D(z^2)$.
Thus, the radial equation, in its linear form, reads as 
\begin{equation}
\partial ^2_r\psi _1 + \partial ^2_z\psi _1 =
-k_0^2D(z^2)\psi _1
\, .
\label{ffpsix}
\end{equation}
In the same approximation, 
the linear electron force balance equation 
can be easily recast. In fact, 
neglecting the function $D(z^2)$, as allowed
by the much greater value of $H$ with respect to $\Delta$,
at the lowest order, we get:
\begin{equation}
v_r \equiv -\frac{1}{r_0\bar{\epsilon }}\partial _z\Theta = 
\frac{\eta c^2k_0^2}{4\pi \partial _{r_0}\psi _0}
\psi _1
\, , 
\label{efbloclin}
\end{equation}
We observe that (as in the dipole magnetic profile)
$\partial _{r_0}\psi _0 < 0$.

In order to deal with the
linear equation (\ref{exazeqlocomb}) which directly links the radial velocity $v_{r}$ to the azimuthal Lorentz force $F_\phi$, we are naturally lead to 
set the conditions $\eta = \eta _0/D(z^2)$ and 
$\mD_{0} = (c^2\eta _0\epsilon _0)/(4\pi)$
($\eta _0$ being the resistivity at $r_0$), obtaining
\begin{equation}
v_r = \frac{1}{8\pi r_0^2\omega _\mathrm{K}\epsilon _0}
\partial _zI\partial _{r_0}\psi_0
\, .
\label{exazeqlocomblin}
\end{equation}
Comparing the two expressions  for $v_r$ (\ref{efbloclin}) and (\ref{exazeqlocomblin}),
we get the compatibility relation
\begin{equation}
\partial _zI =
\frac{2\eta _0\epsilon _0c^2(k_0r_0)^2\omega _\mathrm{K}}
{(\partial _{r_0}\psi _0)^2}
\psi _1
\, , 
\label{exazeqlocombmix}
\end{equation}
Equation (\ref{exazeqlocombmix}) provides a relation between the two functions
$I$ and $\psi _1$. By (\ref{efbloclin}) we easily get
\begin{equation}
\Theta =
- \frac{\eta _0\epsilon _0c^2k_0r_0}{4\pi }
\int dzY 
\, . 
\label{thetafunl}
\end{equation}
Hence, we can find out the vertical and radial velocity components as
\begin{equation}
v_z \simeq \frac{1}{r_0\epsilon _0}
\partial _r\Theta =
-\frac{\eta _0c^2k_0}{4\pi }
\int dz(\partial _rY)
\, , 
\label{v-z}
\end{equation}
\begin{equation}
v_r \simeq -\frac{1}{r_0\epsilon _0}
\partial _z\Theta =
\frac{\eta _0c^2k_0}{4\pi }Y
\, , 
\label{v-rapp}
\end{equation}
where we neglected the function $D(z^2)$ because
its contribution is here of higher order.

Finally. we establish some 
phenomenological relations to shed light on the
physical implications of the proportionality between
the parameters $\mD_{0}$ and $\eta _0$.
According to the analysis developed in Section \ref{sec:3}, the turbulent viscosity coefficient can be expressed as follows
\begin{equation}
\mD_{0} \equiv \frac{2}{3} \alpha \epsilon _0
{v_\mathrm{s}}_0 H
\, ,
\label{formeta}
\end{equation}
where ${v_\mathrm{s}}_0$ denotes the sound velocity
on the equatorial plane.
A reliable estimation for the resistivity
is provided by the relation
\begin{equation}
\eta _0\sim \frac{ m_\mathrm{e}\nu_\mathrm{ie}}{n_\mathrm{e}e^2}
\, ,
\label{revex}
\end{equation}
where $m_\mathrm{e}$ is the electron mass, $n_\mathrm{e}$ the electron
number density and $\nu_\mathrm{ie}$ a typical collision
frequency between ions and electrons.
Comparing equations (\ref{formeta}) and (\ref{revex}) with the relation $\mD_{0} = c^2\eta _0\epsilon _0/(4\pi)$, we can obtain the following estimation for $H$
\begin{equation}\label{nuovastima}
H = \displaystyle\frac{3}{8\pi} \displaystyle\frac{c^{2}m_{e} \nu_\mathrm{ie}}{\alpha n_\mathrm{e} e^{2} v_\mathrm{s0}} = \displaystyle\frac{3}{8\pi}\displaystyle\frac{\nu_\mathrm{ie}}{\alpha n_\mathrm{e} r_\mathrm{c} v_\mathrm{s0}}
\end{equation}
where $r_\mathrm{c}$ is the classical radius of the electron. Equation \reff{nuovastima} establishes a direct relation between the half depth of the disk and the collision frequency between electron and ions once the features of the background quantities $n_\mathrm{e}$ and  $v_\mathrm{s0}$ are assigned.









\section{Non-linear Configuration \label{sec:nonlinearapprox}}

In the non-linear regime, we have seen that the
radial and vertical configurations (\ref{vertad}-\ref{radad}) are described by the
system in the unknowns $Y$ and $\hat{P}$, in
correspondence to assigned $D$ and $\hat{D}$.
In such a non-linear case, we retain also  the direct
relation between the turbulent viscosity coefficient and
the resistivity one, which now reads
\begin{equation}
\eta = 
\frac{4\pi \mD_{0}}{c^2(\bar{\epsilon}
+ \hat{\epsilon})} =
\frac{8\pi \alpha {v_\mathrm{s}}_0H}{c^2(D + \beta ^{-1} \hat{D})} 
\, ,
\label{fundcond}
\end{equation}
reducing \eref{exazeqlocomb} to the
searched link connecting $v_r$ and $F_{\phi}$
\begin{equation}
v_r = \frac{1}{2\omega _\mathrm{K}\epsilon _0(D + \beta ^{-1}\hat{D})} 
F_{\phi }
\, .
\label{exazeqlocombn}
\end{equation}
Substituting $v_r$ by this equation into the
electron force balance \eref{efbloc}, we get the following
expression for $v_z$
\begin{equation}
v_z = \frac{1}{\partial _z\psi _1(D + \beta ^{-1}\hat{D})}     
\left[ \frac{\mD_{0}}{\epsilon _0}
\left(\partial ^2_r\psi _1 + \partial_z^2\psi _1\right)
- \frac{1}{2\omega _\mathrm{K}\epsilon _0}
\left( \partial _{r_0}\psi _0 + \partial _r\psi _1\right) 
F_{\phi } \right]
\, .
\label{efblocn}
\end{equation}
We recall that, when addressing the splitting
$\psi = \psi _0 + \psi _1$, the azimuthal Lorentz
force reads
\begin{equation}
F_{\phi} \simeq 
\frac{1}{4\pi r_0^2}
\left[\partial _zI\left( \partial _{r_0}\psi_0
+ \partial _r\psi _1\right) 
- \partial_rI\partial _z\psi_1\right]
\, .
\label{f0hispl}
\end{equation}
In terms of the function $\Theta$, the expressions
above for $v_r$ (\ref{exazeqlocombn}) and $v_z$ (\ref{efblocn}) stand as
\begin{eqnarray}
\label{thetaformofv}
&\partial _z\Theta = -\displaystyle\frac{r_0}{2\omega _\mathrm{K}} 
F_{\phi }\\
&\partial _r\Theta = \displaystyle\frac{r_0}{\partial _z\psi _1}
\left[ \mD_{0}
\left(\partial ^2_r\psi _1 + \partial_z^2\psi _1\right)
- \frac{1}{2\omega _\mathrm{K}}
\left( \partial _{r_0}\psi _0 + \partial _r\psi _1\right) 
F_{\phi } \right]
\, .
\end{eqnarray}
The solution of the first equation can be taken as
\begin{equation}
\Theta = -\frac{r_0}{2\omega _\mathrm{K}}
\int dz(F_{\phi })
\, ,
\label{thetafuncnl}
\end{equation}
which, substituted in the second, yields an
integro-differential relation for $F_{\phi }$, i.e.
\begin{equation}
\partial _z\psi _1
\int dz(\partial _rF_{\phi }) =
-2\omega _\mathrm{K}\mD_{0}\left(\partial ^2_r\psi _1 + \partial_z^2\psi _1\right)
+ \left( \partial _{r_0}\psi _0 + \partial _r\psi _1\right) 
F_{\phi }
\, .
\label{fphicomp}
\end{equation}
Once the behavior of $\psi _1$ is provided, i.e.
$Y(x, \, u)$, from the equation above we can determine
the form of $F_{\phi }(r,\, z)$ and eventually
calculate the $\phi$-component $I(\psi ,\, z)$ of the magnetic field. It is immediate to check that the
integro-differential equation above provides,
in the linear regime,
the right compatibility condition we
previously fixed in this limit, i.e. eq. (\ref{exazeqlocombmix}).

Equation (\ref{fphicomp}) can be easily rewritten in
the dimensionless form
\begin{equation}
\partial _uY
\int du(\partial _xA_{\phi }) =
-\left(\partial ^2_{x}Y + \epsilon _z
\partial_{u}^2Y\right)
+ \left( 1 + \partial _xY\right) A_{\phi }
\, , 
\label{fphicompdim}
\end{equation}
$A_{\phi }\equiv F_{\phi }/2\omega _\mathrm{K}\mD_{0}k_0$
being a dimensionless function.

\subsection{Solution in the limit $\epsilon _z\rightarrow 0$}

\begin{figure}[ht]
   \centering
   \includegraphics[width=\textwidth]{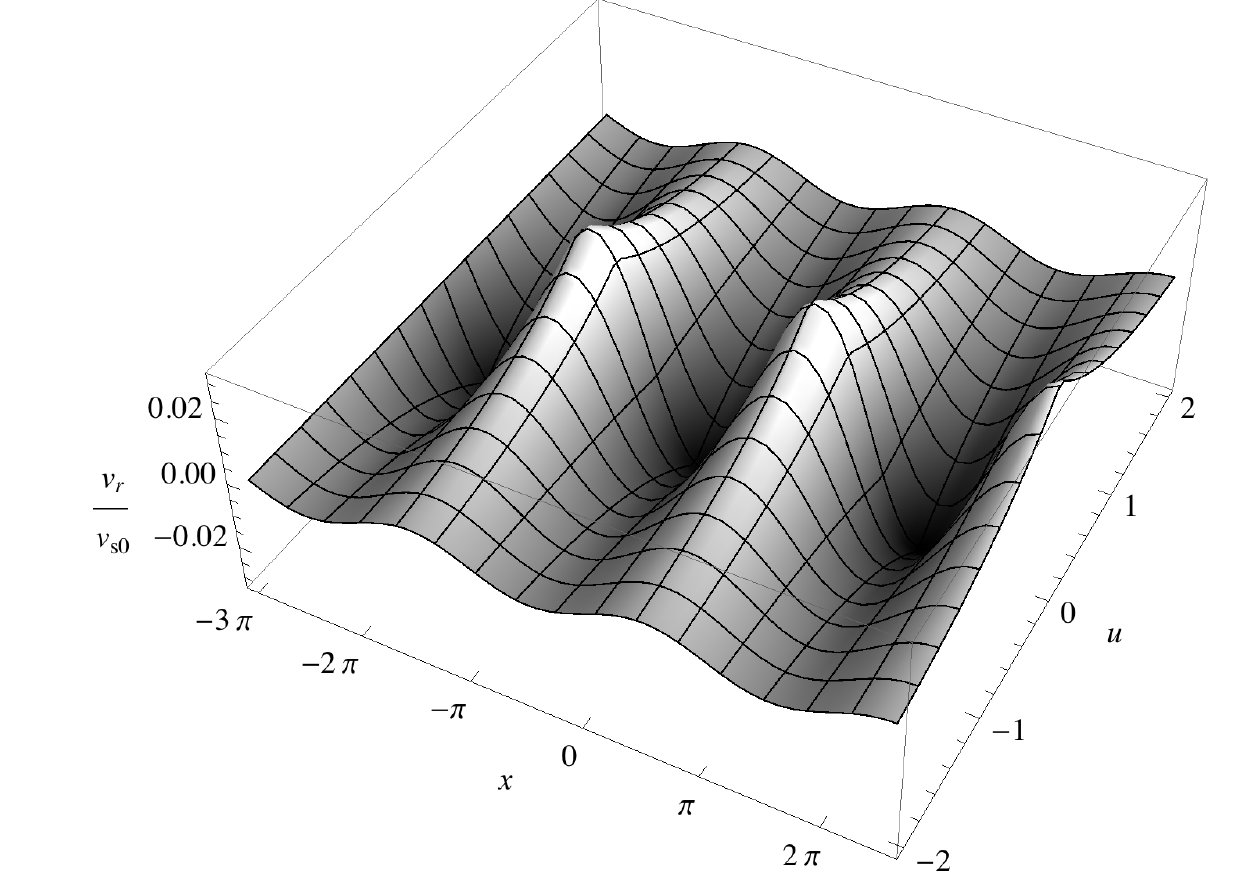} 
   \caption{We plot the radial component of the velocity $v_{r}$ normalized to the constant sound velocity $v_{\mathrm{s}0}$ in correspondence to the value $\alpha=0.01, k_{0} H = 5, \mathcal{A}=1$.}
   \label{fig:velocita}
\end{figure}

The system constituted by the vertical (\ref{vertad})
and the radial (\ref{radad}) equations admits a simple
solution in the limit of vanishing $\epsilon _z$,
as outlined in \citet{C05}.
In fact, such an asymptotic regime can be
consistently represented by the following
expressions for the magnetic surface $Y$ and
the pressure term $\hat{P}$ respectively
\begin{equation}
Y = F(u^2) \sin x \quad , \quad
\hat{P} = Y^2 = F^2(u^2) \sin ^2 x
\, ,
\label{sollim}
\end{equation}
where $F(u^2)$ is a generic function of the vertical
coordinate.
In \citet{C05} this function has been fixed by a
supplementary condition, related with the higher
order in $\epsilon _z$.
Now we show that in the present analysis $F(u^2)$
is naturally fixed by the azimuthal equilibrium,
in particular by the partial integro-differential
\eref{fphicompdim}.
It is immediate to check that if we set
$A_{\phi }(u^2, \, x) = A(u^2)\sin x$, such an
equation can be split into the two relations
\begin{eqnarray}
\label{tworele}
A_{\phi} = - Y\\
\displaystyle\frac{d F}{d u}\int duF = F^2
\label{tworele2}
\end{eqnarray}
The solution of the ordinary integro-differential
equation (\ref{tworele2}) can be taken in the form
\begin{equation}
F\left(u^{2}\right) = \mathcal{A} \exp \left\{ - \sqrt{ u^2}\right\}
\, ,
\label{finsol}
\end{equation}
where $\mathcal{A}$ denotes a positive constant quantity. Thus, by Eq. (\ref{exazeqlocombn}), we get the following
profile of the radial matter infall
\begin{equation}
v_r = -\mathcal{A} \displaystyle\frac{\mD_{0} k_0}{\epsilon _0}
\exp \left\{ - \sqrt{ u^2}\right\}\sin x = \mathcal{A} \displaystyle\frac{2\alpha {v_\mathrm{s}}_0k_0H}{3}
\exp \left\{ - \sqrt{ u^2}\right\}\sin x 
\, ,
\label{radfineq}
\end{equation}
where we made use of the Shakura expression (\ref{formeta})
for the viscosity coefficient $\mD_{0}$. The behaviour of the radial component of the velocity is depicted in fig.\ref{fig:velocita}

\begin{figure}[ht]
   \centering
   \includegraphics[width=.7\textwidth]{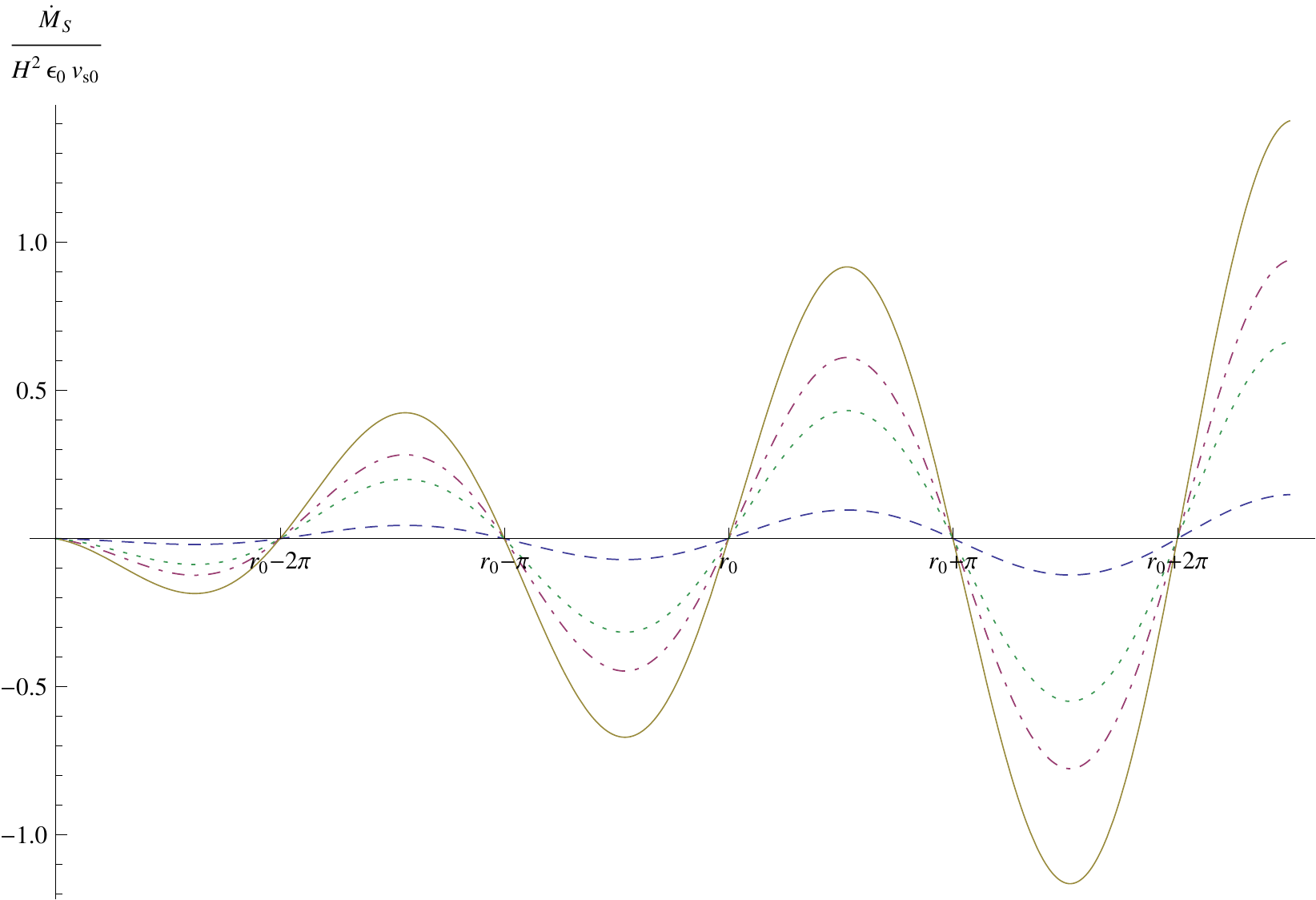} 
   \caption{The behaviour of the accretion rate as a function of the radial coordinate is depicted. It is outlined how the material in-falling from outside in $r_{0}$ is greater than the lost one. The values of the parameters for the dotted, dot-dashed, dashed and thin curve are given by $\alpha = 0.02, 0.02, 0.01, 0.03$ and $\epsilon_{z} = 0.005, 0.01, 0.001, 0.01$ respectively.}
   \label{fig:accrescimento}
\end{figure}

We see how the azimuthal equation completely fixes
the disk profile at the zeroth order in $\epsilon _z$.
The resulting accretion radial velocity has an
oscillatory character along the radial coordinate
around $r_0$, of the same kind found in \citet{MB09}.
According with the discussion pursued in that work,
this velocity profile is able to account for the
local matter infall, but it does not clarify how
the central object can receive this accretion rate.
In fact the following expression for the accretion rate $\dot{M}_{\mathrm{S}}$ holds
\begin{equation}
\displaystyle\frac{{\dot{M}_{\mathrm{S}}}}{\epsilon_{0} H^{2} v_{s0}} = \displaystyle\frac{8\pi}{3} \mathcal{A} \alpha\sqrt{\epsilon_{z}} (x+k_{0}r_{0}) \sin x\,,
\end{equation}
leading to the non zero average for $x\in [-\pi,\pi]$
\begin{equation}
\left\langle\displaystyle\frac{\dot{M}_{\mathrm{S}}}{\epsilon_{0} H^{2} v_{s0}}\right\rangle = \displaystyle\frac{16\pi^{2}}{3} \mathcal{A} \alpha\sqrt{\epsilon_{z}}\,.
\end{equation}
This behaviour is sketched in fig.\ref{fig:accrescimento}

The present analysis, differently from the linear
case addressed in \citet{MB09}, lives in a full
non-linear regime $Y\sim 1$ (i.e. $\mathcal{A}\sim 1$).
Therefore we are able to identify the mass perturbation
$\hat{\epsilon}$, as the relevant
feature to deal with a global matter infall.
Indeed, it seems that the non-linearity of the
equilibrium configuration is not a  strong enough feature to
excite modes with non-zero radial averaged
infall velocity.

\begin{figure}[ht]
   \centering
   \includegraphics[width=.6\textwidth]{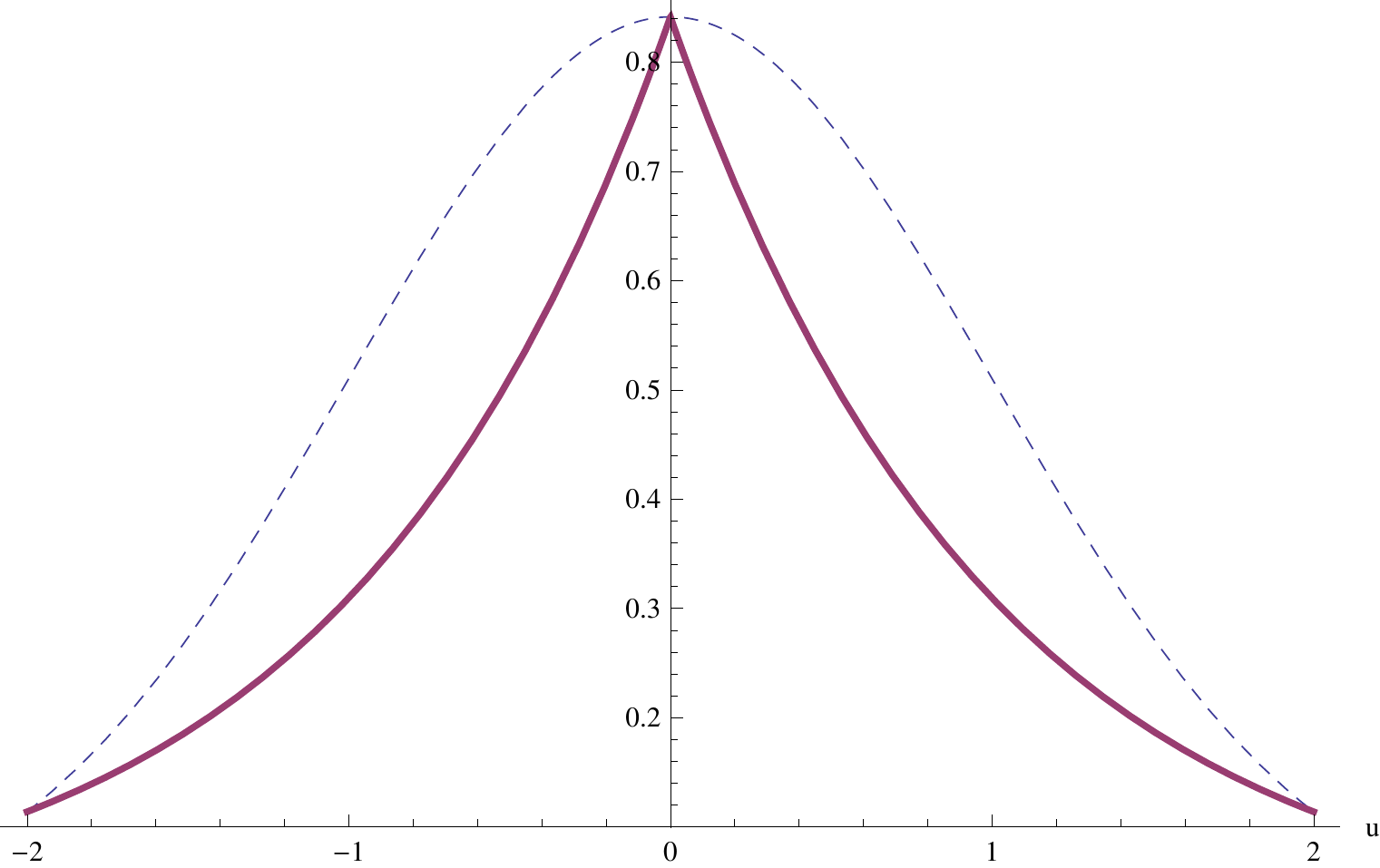}
   \caption{The vertical dependence of the disk profile $F(u^{2})$ is plotted in the present case (thin line) in comparison to the ideal MHD case (dashed line).}
   \label{fig:confronto}
\end{figure}

Finally we observe how the vertical decay of the
magnetic flux surface, over the equatorial plane,
is slower in the present case with respect to the
analysis pursued in \citet{C05}, as outlined in Fig\ref{fig:confronto}.

\begin{figure}[ht]
   \centering
	   \includegraphics{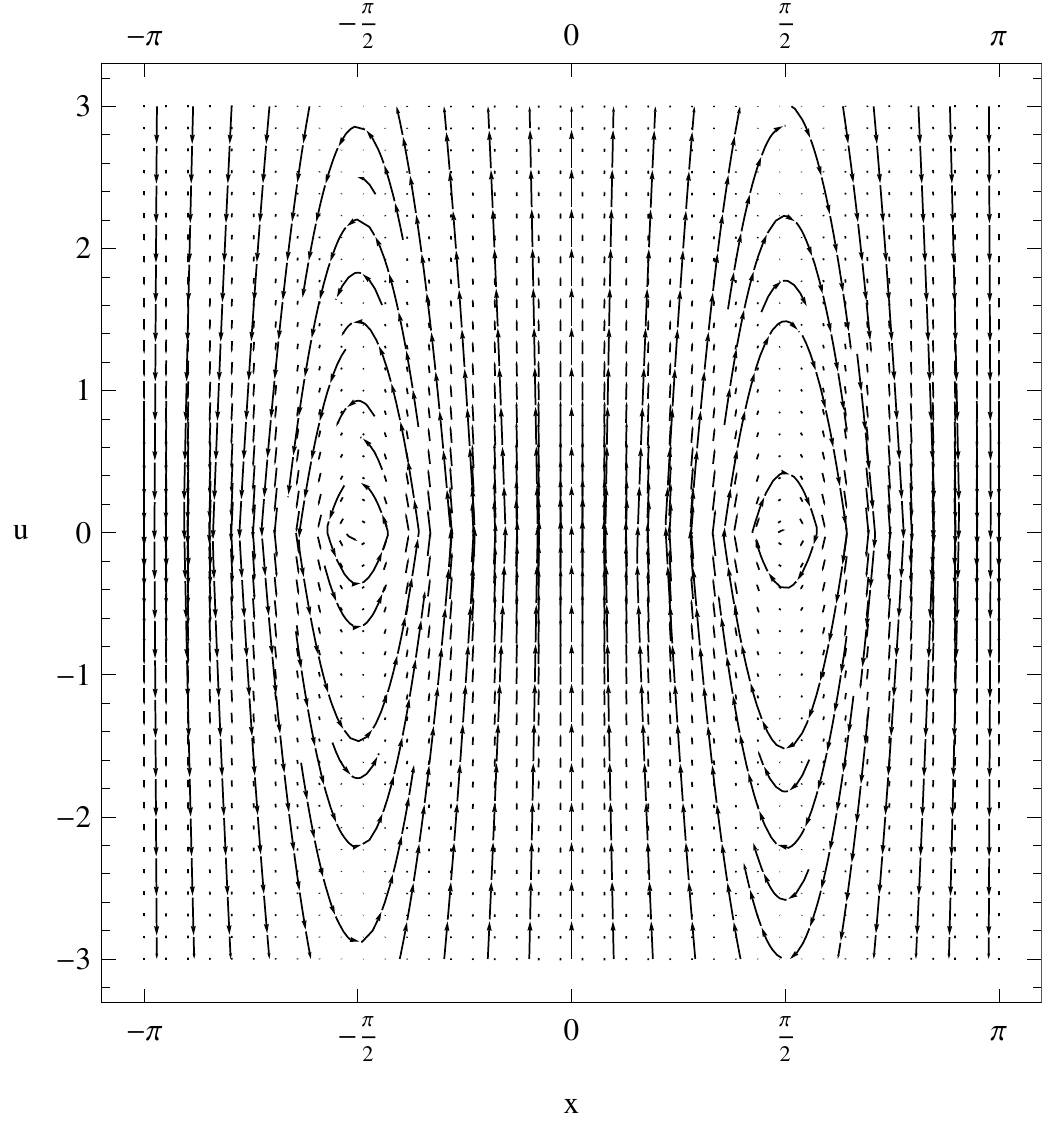} 
   \caption{The magnetic field lines neglecting the toroidal components, in correspondence to the values $\mathcal{A} = 1, \sqrt{\epsilon_{z}} =0.005$}
   \label{fig:Bfield}
\end{figure}

By other words, accounting for the azimuthal and electron
force balance equations allows a higher intensity
of the magnetic field in a given vertical region. The form of the total magnetic surfaces corresponds to the expression
\begin{equation}
Y_{\mathrm{T}}(x, u^{2}) = x + Y(x, u^{2}) = x +\mathcal{A} \exp\left(-\sqrt{u^{2}}\right) \sin(x)\,.
\end{equation}
Neglecting the toroidal component, which is regarded as small, the magnetic field components take the form
\begin{subequations}
\begin{equation}
\displaystyle\frac{B_{r}}{B_{z0}} = -\displaystyle\sqrt{\epsilon_{z}} \partial_{u} Y_{\mathrm{T}} = \displaystyle\sqrt{\epsilon_{z}} \mathcal{A} \frac{u}{\sqrt{u^{2}}} \exp\left(-\sqrt{u^{2}}\right) \sin(x)\,,
\end{equation}
\begin{equation}
\displaystyle\frac{B_{z}}{B_{z0}} =  \partial_{x} Y_{\mathrm{T}} = 1 + \mathcal{A} \exp\left(-\sqrt{u^{2}}\right) \cos(x)\,,
\end{equation}
\end{subequations}
The $\vec{B}$-field line are depicted in Fig.\ref{fig:Bfield}.

\section{Concluding Remarks}

Our analysis constitutes a bridge scheme between the standard one-dimensional model for an accretion disk \citep{B01} and the reformulation provided in \citet{C05} in terms of a local bidimensional MHD profile of the magnetically confined plasma. The original description relies on the setting of a viscoresistive MHD framework in which the angular momentum transport is allowed by the turbulent viscosity coefficient arising when the Reynolds number of the fluid has sufficiently high value. 
This paradigm has been able to provide with a satisfactory interpretation of some observed feature in strongly accreting system like X-ray binary pulsars.

On the other hand, the approach introduced in \citet{C05} starts from a criticism concerning the possibility to generate in the plasma of the disk significant viscosity and resistivity effects. Therefore the disk morphology has been addressed on a purely theoretical point of view, relying on the hypotheses of the ideal bidimensional MHD  in axial symmetry. However this point of view cannot properly account for the accretion mechanism because it stands only for a velocity field completely dominated by the Keplerian rotation. In fact, including radial and vertical velocity in this bidimensional MHD scheme requires a proper setting of the electron force balance equation.

We have constructed a scheme in which the viscoresistive bidimensional MHD scenario is implemented in the same spirit of \citet{C05}, but with the advantage to be able to account for the azimuthal and electron force balance equilibria, necessary to fix the accretion properties of the system. The resulting approach emerged as self consistent and, as far as a direct relation between the viscosity and the resistivity coefficients was addressed, we could fix the radial and vertical profile of the magnetic flux surfaces, due to the electromagnetic back-reaction. We devoted particular attention to the analysis of the linear case when the induced magnetic surface is a small correction only to the background features. Furthermore we solved the non-linear case in the limit of a vanishing ratio between the magnetic pressure to the thermostatic one.

The results we derived above constitute the first step for a reformulation of the mechanism at the ground of the accretion phenomenon in terms of local structures arising in the disk morphology, instead of the diffusive description characterizing the standard model.

This  work  was developed within the framework of the CGW Collaboration \\\textsf{(www.cgwcollaboration.it)}.

%

\begin{thebibliography}{16}
\providecommand{\natexlab}[1]{#1}
\providecommand{\url}[1]{{#1}}
\providecommand{\urlprefix}{URL }
\expandafter\ifx\csname urlstyle\endcsname\relax
  \providecommand{\doi}[1]{DOI~\discretionary{}{}{}#1}\else
  \providecommand{\doi}{DOI~\discretionary{}{}{}\begingroup
  \urlstyle{rm}\Url}\fi
\providecommand{\eprint}[2][]{\url{#2}}

\bibitem[{{Balbus} and {Hawley}(1998)}]{B98}
{Balbus} SA, {Hawley} JF (1998) Instability, turbulence, and enhanced transport
  in accretion disks. Rev Mod Phys 70:1, \doi{10.1103/RevModPhys.70.1}
 
\bibitem[{{Benini} and {Montani}(2009)}]{MGXII}
{Benini} R, {Montani} G (2009) {2-D MHD Configurations for Accretion Disks
  Around Magnetized Stars}. ArXiv e-prints \eprint{0912.1721}

\bibitem[{Bisnovatyi-Kogan and Lovelace(2001)}]{B01}
Bisnovatyi-Kogan GS, Lovelace RVE (2001) Advective accretion disks and related
  problems including magnetic fields. New Astro Rev 45:663,
  \doi{10.1016/S1387-6473(01)00146-4}

\bibitem[{{Coppi}(2005)}]{C05}
{Coppi} B (2005) ``crystal'' magnetic structure in axisymmetric plasma
  accretion disks. Physics of Plasmas 12:7302, \doi{10.1063/1.1883667}

\bibitem[{Coppi and Rousseau(2006)}]{CR06}
Coppi B, Rousseau F (2006) Plasma disks and rings with ``high'' magnetic energy
  densities. ApJ 641:458, \doi{10.1086/500315}

\bibitem[{Ferraro(1937)}]{Ferraro}
Ferraro VCA (1937) The non-uniform rotation of the sun and its magnetic field.
  Mon Not R Astron Soc 97:458

\bibitem[{{Lynden-Bell}(1996)}]{L96}
{Lynden-Bell} D (1996) Magnetic collimation by accretion discs of quasars and
  stars. Mon Not R Astron Soc 279:389

\bibitem[{Lynden-Bell and Pringle(1974)}]{lyndenpringle74}
Lynden-Bell D, Pringle JE (1974) The evolution of viscous discs and the origin
  of the nebular variables. Mon Not R Astron Soc 168:603

\bibitem[{{Montani} and {Benini}(2009)}]{MB09}
{Montani} G, {Benini} R (2009) {Linear Two-Dimensional MHD of Accretion Disks:.
  Crystalline Structure and Nernst Coefficient}. Modern Physics Letters A
  24:2667--2680, \doi{10.1142/S0217732309031879}, \eprint{0909.0371}

\bibitem[{{Ogilvie}(1997)}]{Ogilvie97}
{Ogilvie} GI (1997) {The equilibrium of a differentially rotating disc
  containing a poloidal magnetic field}. Mon Not R Astron Soc 288:63--77

\bibitem[{Pringle and Rees(1972)}]{pringlerees72}
Pringle JE, Rees MJ (1972) Accretion disc models for compact x-ray sources.
  Astron Astrophys 21:1

\bibitem[{{Ruffini} and {Wilson}(1975)}]{RuffiniWilson}
{Ruffini} R, {Wilson} JR (1975) {Relativistic magnetohydrodynamical effects of
  plasma accreting into a black hole}. Phys Rev D 12:2959--2962,
  \doi{10.1103/PhysRevD.12.2959}

\bibitem[{Shakura(1973)}]{S73}
Shakura NI (1973) Disk model of gas accretion on a relativistic star in a close
  binary system. Soviet Astronomy 16:756

\bibitem[{Spruit(2008)}]{Sp08}
Spruit H (2008) Theory of magnetically powered jets. arXiv 0804:3096
 
\bibitem[{Velikhov(1959)}]{V59}
Velikhov E (1959) {Stability of an ideally conducting liquid flowing between
  cylinders rotating in a magnetic field}. Sov Phys JETP 36(9):995--998

\bibitem[{Verbunt(1982)}]{verbunt82}
Verbunt F (1982) Accretion disks in stellar x-ray sources - a review of the
  basic theory of accretion disks and its problems. Space Science Reviews
  32:379, \doi{10.1007/BF00177448}
 

\end{thebibliography}

\end{document}